\documentstyle[12pt,axodraw]{article} 
\begin{document}
\pagestyle{empty}
\begin{flushright}
   UICHEP-TH/99-8\\[-0.1cm]
   hep-ph/9910465\\[-0.1cm]
   September 1999
\end{flushright}

\begin{center}
{\Large {\bf Additional two--loop contributions to electric dipole moments
in supersymmetric theories}}\\[1.cm]
{\large 
Darwin Chang$^{a,b}$, We-Fu Chang$^{a,b}$ and Wai-Yee Keung$^{c,b}$}\\

{\em $^a$NCTS and Physics Department, National Tsing-Hua University,\\
Hsinchu 30043, Taiwan, R.O.C.}\\
{\em $^b$Fermilab, P.O. Box 500, Batavia IL 60510, U.S.A.}\\
{\em $^c$Physics Department, University of Illinois at Chicago, IL
  60607-7059,  USA}\\
\end{center}
\bigskip\bigskip
\centerline{\bf  ABSTRACT} 
We calculate the two--loop contributions to the electric dipole moments of
the electron and the neutron mediated by charged Higgs in a generic
supersymmetric theories.
The new contributions are originated from the  potential CP violation in
the trilinear couplings of the charged Higgs bosons to the scalar-top or
the scalar-bottom quarks.  
These couplings did not receive stringent constraints directly.
We find observable effects for a sizeable portion of the parameter space
related to the third generation scalar-quarks in the minimal
supersymmetric standard model.\\[0.3cm]

\newpage
\pagestyle{plain}

It is widely believed theoretically that supersymmetry will play a
significant role at certain fundamental scale that can be probed in the
near future.  As a result, it is one of the major goal of the new and 
the future colliders to look for signature of supersymmetry.

The strongest constraint on supersymmetry is from the flavor changing
neutral current process such as $K -\bar{K}$ mixing and $\mu \rightarrow e
\gamma$, or from the flavor neutral CP violating quantities such as the
electric dipole moments (EDM's) of the electron and the neutron\cite{masiero}.
These quantities
put strong limits on  new parameters related to the first two
generations of quarks and leptons in any generic supersymmetric theories
resulting in the famous SUSY flavor problem and the SUSY CP problem.  A
solution will require theory to explain why these parameters are so small.
One scenario out of these problems is to assume that the first two
generations of squarks and sleptons are heavy, while keeping the third
generation relatively light in order to produce interesting physical
consequence such as low energy electroweak baryogenesis\cite{ewbg}.
Such scenario is consistent with the sparticle spectrum resulting from
renormalization group evolution starting from an high energy unified
theory.  It is therefore interesting to look for effective constraint on
the new SUSY parameters related to the third generation using the low
energy data or collider searches.  

In a recent paper\cite{ckp}, it has been shown that by considering the
two--loop contributions to the electric dipole moment of the electron and
the neutron, it is possible to put nontrivial constraint on the CP
violating parameters related to the third generation directly.  In
fact it was shown later\cite{baek} that such constraint actually helps
rule out a scenario of the supersymmetric theory in which the CP
violating $\epsilon$ parameter in the kaon system was proposed to be
originated purely from a new  source beyond the standard model.\cite{demir}.

In Ref.\cite{ckp}, only the two--loop diagrams with the second loop 
mediated by a neutral Higgs boson have been considered.  
These diagrams are of Barr-Zee type\cite{barrzee}.  However, there are
scenarios in supersymmetric theories in which other scalar bosons, such as
the charged Higgs boson, may be light enough to produce larger effect.
In this paper we wish to extend the earlier analysis to include
two--loop contributions with the second loop involving the exchange of a
charged Higgs boson.  It is possible to analyse the two sets of two--loop
contributions separately because they are gauge invariant independently.  The
charged Higgs set is more technically sophisticated and has only been
studied in non-supersymmetric theories in Ref.\cite{bck}.  
\begin{center}
\begin{picture}(200,150)(0,10)
\ArrowLine(10,30)(33,47)\Text(25,25)[lb]{$\ell_R$}
\DashArrowLine(33,47)(68,85){5}
\Text(55,75)[rb]{\small $H^-$}
\Photon(92,85)(127,47){2}{6}
\Text(103,80)[l]{\small $\searrow W^-,q,\alpha$}
\ArrowLine(33,47)(127,47)
\ArrowLine(127,47)(155,30)\Text(120,25)[l]{$\ell_L$}
\Photon(80,120)(80,150){3}{3}
\Text(86,140)[l]{$\uparrow\gamma,k,\mu$}
\GOval(80,100)(20,20)(0){.8}
\Text(80,100)[b]{$\Gamma^{\mu\alpha}$}
\end{picture}
\\
Fig.~1.  Structure of the two--loop EDM diagram via Barr-Zee mechanism
in supersymmetric theories with charged Higgs exchange in the second loop.
\end{center}
The two--loop diagrams are shown as in Fig.~1, in which both the stop and
the sbottom are involved in the first shaded loop.  
In a generic gauge, there are
also other diagrams involving the unphysical Higgs boson $G^\pm$.
However it is known\cite{gavela}
that in  $R_\xi$  {\it nonlinear} Landau gauge (NLLG),
such diagrams will not contribute.  For this reason, our calculation
is based on NLLG.

We shall first discuss the first loop 
that creates an $H\gamma W^*$ vertex and then integrate the second loop
in the EDM calculation. 

\section*{The first loop of $H\gamma W^*$ Vertex}
The relevant interaction Lagrangian in our study is given by,
\begin{equation} 
{\cal L} =  v  H^{-*}\tilde t^* {\cal E} \tilde b
         + {g_2\tan\beta m_\ell \over\sqrt2 M_W}  \bar \nu_L \ell_R H^{-*}
         -\hbox{$g_2\over\surd2$}W_\sigma^{-*}
(\bar\ell_L\gamma^\sigma\nu_L + \tilde t^* {\cal K}
i{\stackrel{\leftrightarrow}{\partial^\sigma}} \tilde b) 
   \hbox{ + H.c. }  
\label{eq:mainL}
\end{equation}
plus mass terms.  
Here $\tilde t$ and $\tilde b$ carry indices $1,2$  labeling the
corresponding mass eigen-states which are superpositions of the
bosonic chirality states,
\begin{equation}\left(
     \begin{array}{c} \tilde b_L \\  \tilde b_R\\ \end{array}
   \right)
= {\large U_b}
    \left(
     \begin{array}{c} \tilde b_1 \\  \tilde b_2\\ \end{array}
    \right)  \ ,\quad
\left(
     \begin{array}{c} \tilde t_L \\  \tilde t_R\\ \end{array}
   \right)
= {\large U_t}
    \left(
     \begin{array}{c} \tilde t_1 \\  \tilde t_2\\ \end{array}
    \right)          \ . \end{equation}
These unitary transformation $U_t$ and $U_b$ 
diagonalize the mass matrices 
\begin{equation}
{\cal M}^2_{\tilde{q}} =
\left( \begin{array}{cc}
M^2_{\tilde q_L}  + m_q^2 +\Delta_{q_L} M_Z^2\cos2\beta & 
- (\hat m_q\mu    +m_q   A_q^*)\\ 
- (\hat m_q\mu^*  +m_q   A_q)& 
M^2_{\tilde q_R} + m_q^2+\Delta_{q_R} M_Z^2\cos2\beta\\
\end{array}\right) \ ,
\label{eq:Mass}
\end{equation}
with $q=t,b$  with $\hat m_b=m_b\tan\beta$, $\hat m_t=m_t\cot\beta$, 
$\Delta_{q}=T^3_q-e_q\sin^2\theta_W$ and 
the $SU(2)$ relation $M_{\tilde t_L} = M_{\tilde b_L}$.
The complex phase can be factored out as
\begin{equation} 
U_q=\left(\begin{array}{cc} 1 & 0 \\ 0 & e^{i\delta_q}\end{array}\right)
       \left(\begin{array}{cc} c_q & s_q \\ 
                              -s_q & c_q \end{array}\right) \ ,
\label{eq:U}
\end{equation}
with  $c_q\equiv \cos\theta_q$,  $s_q\equiv \sin\theta_q$ for $q=b,t$,
and
\begin{equation}
\hat m_q\mu^*  +m_q   A_q=|\hat m_q\mu^*  +m_q   A_q|e^{i\delta_q} 
\ .\end{equation}

The charged current coupling matrix ${\cal K}$ in this basis of the 
$\tilde t$--$\tilde b$  sector is {\it real},
\begin{equation}
\left({\cal K}\right)={\large U^\dagger_t}
 \left(\begin{array}{cc} 1 & 0 \\ 0 & 0  \end{array} \right) {\large U_b}
=\left(\begin{array}{cc} c_tc_b & c_ts_b \\
                         s_tc_b & s_ts_b \end{array} \right) \  .\end{equation}
The tri-linear scalar terms for charged Higgs boson can be written as
\begin{equation}
\left({\cal E}\right)={\surd2\over v^2}
{\large U^\dagger_t}
 \left(\begin{array}{cc}   
-M_W^2\sin2\beta+m_b^2\tan\beta+m_t^2\cot\beta & m_b\mu-A_b^*\hat m_b \\
m_t\mu^*-A_t\hat m_t &    2m_b m_t/\sin2\beta \\ \end{array} 
\right)
                    {\large U_b} \ .
\label{eq:E}
\end{equation}
The imaginary entries are CP violating,
\begin{eqnarray}
\hbox{Im}\left({\cal E}\right) &=&
{\sqrt{2}\over v^2} {2m_b m_t\over \sin2\beta}
\sin(\delta_b-\delta_t) \left(\begin{array}{cc}   
s_t s_b  & -s_t c_b \\
-c_t s_b & c_t c_b \end{array}\right)  \nonumber
\\ && \quad +
\left( \begin{array}{cr}
-\, c_t s_b\, \lambda_b\, +\,
s_t c_b \, \lambda_t & \quad
c_t c_b\, \lambda_b\, +\,
s_t s_b\, \lambda_t\\
-\,s_t s_b\, \lambda_b\, -\,
c_t c_b\, \lambda_t &
s_t c_b\, \lambda_b\, -\,
c_t s_b\, \lambda_t \end{array} 
\right)
\ . 
\label{eq:Im}
\end{eqnarray}
where 
\begin{equation}
\lambda_q = {\sqrt{2} \hat m_q\over v^2 \sin\beta\cos\beta} 
                             {\rm Im}\left(\mu e^{i\delta_q}\right)
\ .
\label{eq:lambda}
\end{equation}
CP violation is a result of the mismatch in the phase between the
${\cal E}$ vertex of the charged Higgs boson and the ${\cal K}$ vertex
of the $W$ boson.  Our convention happens to give rise to real ${\cal
K}$ and complex ${\cal E}$.  Note that while there are two stop and
two sbottom eigenstates, we actually need only one of them each to
give contribution to the EDM of leptons or quarks.  Therefore, one can
in principle consider the numerically simpler limit in which one of
the squark of each flavor is much lighter than the other one.  Also,
even if the squark mass matrices are accidentally real, the CP
violation can still be generated through the 
weak-basis-charged-Higgs-coupling matrix in Eq.(\ref{eq:E}).  
In the first term of Eq.~(\ref{eq:Im}), 
the CP violation is due to the relative phase
between stop and sbottom mass matrices, while in the second term CP
violation is due to the relative phase between the 
weak-basis-charged-Higgs-coupling matrix and squark mass matrices.
Note, however, that if $\lambda_t = \lambda_b = 0$, the phase of $\mu$ may
not be zero, however $\sin(\delta_b - \delta_t)$ will be zero and there
will be no CP violation in the charged Higgs interaction in our
approximation (of ignoring the squark flavor mixing).

We work out the 1-loop amplitude for the process
$H(k+q) \to \gamma(k,\mu) + W^*(q,\alpha)$. 
Besides the 3-prong irreducible diagram, Fig.~2(a),
\begin{center}
\begin{picture}(200,110)(10,40)
\DashArrowLine(33,47)(68,85){5}
\Text(55,75)[rb]{\small $H^-$}
\Photon(92,85)(127,47){2}{6}
\Text(103,80)[l]{\small $\searrow W^-,q,\alpha$}
\Photon(80,120)(80,150){3}{3}
\Text(86,140)[l]{$\uparrow\gamma,k,\mu$}
\DashArrowArc(80,100)(20,0,360){3}
\Text(55,100)[r]{\small $\tilde{t},l-k$}
\Text(103,100)[lb]{\small $\uparrow\tilde{t},l$}
\Text(80,87)[b]{\small $\tilde b$}
\Text(80,80)[t]{\small $\stackrel{\rightarrow}{\tiny l+q}$}
\Text(80,35)[b]{(a)} 
\end{picture}
\begin{picture}(120,110)(-10,-50)

\Photon(20,0)(50,30){2}{6} \Text(45,35)[l]{$\mu$}
                           \Text(55,30)[l]{$\nearrow,k,\gamma$}
\Photon(20,0)(50,-30){2}{6}\Text(55,-30)[l]{$\alpha$}
                           \Text(40,-15)[l]{$\searrow q,W^-$}
\Photon(0,0)(20,0){2}{4}   
\DashCArc(-10,0)(10,0,360){3}
\Text(15,-5)[t]{$\small \beta$}
\Text(5,-5)[t]{$\small \sigma$}
\Text(-10,20)[b]{$p\to $}
\DashArrowLine(-50,0)(-20,0){5}\Text(-35,-5)[t]{$H^-$}
\Text(0,-55)[b]{(b)} 
\end{picture}
\begin{picture}(120,110)(-35,-50)
\Photon(20,0)(50,30){2}{6} \Text(45,35)[l]{$\mu$}
                           \Text(55,30)[l]{$\nearrow,k,\gamma$}
\Photon(20,0)(50,-30){2}{6}\Text(55,-30)[l]{$\alpha$}
                           \Text(40,-15)[l]{$\searrow q,W^-$}
\DashCArc(0,0)(20,0,360){3}
\DashArrowLine(-50,0)(-20,0){5}\Text(-35,-5)[t]{$H^-$}
\Text(0,-55)[b]{(c)} 
\end{picture}
\\
Fig.~2 One loop $H^-\to W^-\gamma$ Feynman diagrams in the {\it non-linear}
Landau gauge.
\end{center}
there is  a one-particle reducible $H^-$--$W^-$ bubble diagram, Fig.~2(b), 
with the external photon $k$
attached to the outgoing leg $W^-$. 
This amplitude is not vanishing
even at the limit of the gauge parameter $\xi\to 0$ 
in our calculation based on the $R_\xi$
{\it non-linear} Landau gauge (NLLG), where
the unphysical Goldstone boson coupling in $GW^-\gamma$ is absent.
One notices that the $W$ propagator has the form
\begin{equation}
{\cal P}^{\sigma\beta}(p) =
 -{g^{\sigma\beta} + {p^\sigma p^\beta (\xi-1)\over p^2-\xi M_W^2}
                                       \over p^2-M_W^2}
\stackrel{\xi\to 0}{\longrightarrow}
  -{g^{\sigma\beta} -  p^\sigma p^\beta/p^2    \over p^2- M_W^2}
  -{\xi p^\sigma p^\beta \over (p^2)^2}  \  .\end{equation}
The first term gives vanishing result when it is convoluted with the 
bubble amplitude of the form $Bp_\sigma$. 
The second term will give finite result even when $\xi\to 0$ as there
are singular terms from the tri-gauge-boson coupling
in the Non-Linear Landau gauge,
\begin{equation} \xi(-B/p^2) 
  [(p+k+q/\xi)^\alpha p^\mu
  +(-k+q-p/\xi)\cdot  p g^{\mu\alpha}
  +(-q-p)^\mu p^\alpha] \longrightarrow  B g^{\mu\alpha} \ ,  \end{equation}
where we have extracted the leading constant term in the limit $\xi\to  0$.
We need also to include a sea-gull diagram, Fig.~2(c), with the external 
photon $k$ sticked to the $\tilde t$--$\tilde b$--$W$ vertex.
The photon $(k,\mu)$ is attached to the external electric field for
the EDM measurement. We only keep the lowest order of $k$ in our
calculation.
\begin{equation}
 i\Gamma^{\mu\alpha}=-i\sum_{c,d}
{{\cal E}^*_{c,d}  v g_2 {\cal K}_{c,d} e
\over 8\pi^2\surd2}
\int^1_0{3[e_t(1-y)+e_by] y(1-y)dy 
\over 
(1-y)m_{\tilde t_c}^2+y m_{\tilde b_d}^2-q^2y(1-y)}
(k^\alpha q^\mu-k\cdot q g^{\mu\alpha}) \ .
\label{eq:gWH}
\end{equation}
Note that the amplitude in NLLG satisfies the Ward identity 
$k_\mu\Gamma^{\mu\alpha} =0$  even when $q^2\ne M_W^2$. 
\section*{The second loop and EDM}

The final 2-loop EDM amplitude (Fig.~1) of the lepton becomes 
\begin{eqnarray}
  i{\cal M}_{H^+} &=&i\sum_{c,d} e{\hbox{Im}({\cal E}^*_{cd}){\cal K}_{cd}
m_\ell \tan\beta 
g^3_2 v i\over 512\pi^4 \sqrt2 M_W}
\int {3(e_t(1-y)+e_b y)y(1-y)dy\over 
(1-y)m_{\tilde t_c}^2+ym^2_{\tilde b_d}+Q^2y(1-y)}    \nonumber
\\ &&\qquad \qquad \times 
           {dQ^2\over Q^2+M_{H^-}^2} {Q^2\over Q^2+M_W^2} 
        {i\over 2}\sigma^{\mu\nu}k_\nu\gamma_5 \ .
\end{eqnarray}
Here we integrate $Q^2$ over $(0,\infty)$, and $y$ over $(0,1)$. 
In our EDM convention,
${\cal M}=-\sigma^{\mu\nu} k_\nu \gamma_5 d_\ell$, we have
the contribution to EDM from the charged Higgs sector,
$$\left({d_\ell\over e}\right)_{H^\pm}
=\sum_{c,d}{\hbox{Im}({\cal E}^*_{c,d}) {\cal K}_{c,d} m_\ell \tan\beta
                   g_2^2          \over 512\pi^4\sqrt2}
\int {3[e_t(1-y) + e_b y] {y(1-y) Q^2 dy dQ^2 \over (
Q^2+M_{H^-}^2)(Q^2+M_W^2)}
\over 
(1-y)m_{\tilde t_c}^2+ym_{\tilde b_d}^2+Q^2y(1-y)}  $$
$$
\qquad\quad = -3\sum_{c,d} {\hbox{Im}({\cal E}^*_{c,d}){\cal K}_{c,d} 
                      \alpha_{\rm em} \tan\beta
          \over 256\sqrt{2}\pi^3\sin^2\theta_W }{m_\ell\over M_{H^-}^2}
       \left[ e_t {\cal F}\left({M_W^2\over M_{H^-}^2},
                  {M_{\tilde t_c}^2\over M_{H^-}^2},
                  {M_{\tilde b_d}^2\over M_{H^-}^2}\right)  
\right.$$
\begin{equation} \hbox{\hskip 3.5in}
    +\left. e_b {\cal F}\left({M_W^2\over M_{H^-}^2},
                  {M_{\tilde b_d}^2\over M_{H^-}^2},
                  {M_{\tilde t_c}^2\over M_{H^-}^2}\right)\right] \ ,
\label{eq:edmW}
\end{equation}
\begin{equation} {\cal F}(w,\tau,\beta) =
{1\over 1-w}\left[{\cal F}_0(\tau,\beta)
                 -{\cal F}_0\left({\tau\over w},{\beta\over w}\right)\right]
\ , \end{equation}
\begin{equation}  {\cal F}_0(\tau,\beta)=
\int_0^1 
{2y(1-y)^2 \ln {y(1-y) \over (1-y)\tau+y\beta}
 \over \tau(1-y)+\beta y-y(1-y)} dy
\ .\end{equation}
The above integral is derived based on the relation,
$$\int_0^\infty{xdx\over (x+a)(x+b)(x+c)}
={1\over a-b}\left[{\ln (a/c)\over 1-c/a}
                  -{\ln (b/c)\over 1-c/b}\right]
\ . $$
The value of ${\cal F}_0$ is negative and approaches zero as the arguments
$\tau, \beta$ approach infinity.  Therefore ${\cal F}$ is also nagetive.  
Note that $\sum_{c,d}{\rm Im}({\cal E}^*_{c,d}){\cal K}_{c,d} = 0$ 
as expected because if 
$M_{\tilde q_1} = M_{\tilde q_2} (q = t, b)$ 
the mass matrix ${\cal M}^2_{\tilde{q}}$ is proportional 
to the diagonal matrix and the CP violating phase $\delta_q$ disappears.  
In that case, even though the
charged Higgs coupling matrix in Eq.(\ref{eq:E}) may appear to be complex,
but the off diagonal phases can easily be absorbed into ${\tilde t_R}$ and
${\tilde b_R}$.  Note also that if $M_{\tilde b_1} = M_{\tilde b_2}$, the
CP violation will depends on $\lambda_t$ only and  the contribution from the
first term in Eq.(\ref{eq:Im}) will disappear as expected.

\section*{Comparison with Pseudoscalar Contribution}
The neutral $A^0$ couples to stop and sbottom in the form
\begin{equation}
 {\cal L}_{A^0}=vA^0\left[ \tilde t^\dagger \left(\xi_t\right) \tilde t
                            +\tilde b^\dagger \left(\xi_b\right) \tilde b    
                     \right]
+{g_2 m_\ell\over 2M_W} \tan\beta  A^0\bar \ell (i\gamma_5) \ell
\  . \end{equation}
The real part of the matrix $(\xi)$ is CP violating,
\begin{equation}
\hbox{Re}\left(\xi_q\right)={\lambda_q\over \sqrt2}\sin2\theta_q
\left(\begin{array}{cc}
                      1 & -\cot2\theta_q\\
                    -\cot2\theta_q & -1 \end{array}\right) \ . \end{equation}
We obtain
\begin{equation} 
  \label{EDMf}   
\left( \frac{d_\ell}{e}\right)_{A^0\gamma} \ = -
\frac{3 e_\ell \alpha_{\rm em}}{32\pi^3}\, \frac{m_\ell}{M^2_{A^0}}
 \tan\beta \sum_{q = t,b}\ 
{\lambda_q\over \sqrt2}\sin2\theta_q         e^2_q\,\left[\, 
F\left({M^2_{\tilde{q}_1} \over M^2_{A^0}}\right)
 -
F\left({M^2_{\tilde{q}_2} \over M^2_{A^0}}\right) \right]\, ,
\label{eq:Agamma}
\end{equation}
with 
\begin{equation} F(\tau)={\cal F}(0,\tau,\tau)={\cal F}_0(\tau,\tau)=
 \int_0^{1}  \frac{y(1-y)}{\tau - y(1-y)} \ln {y(1-y)\over \tau} dy \
 . 
\end{equation}
Note that, as expected, if both 
$M_{\tilde q_1} = M_{\tilde q_2}$ for $(q = t, b)$
CP violation disappear in our approximation.  This explains the
cancelation in Eq.(\ref{eq:Agamma}).

Similar contribution from the $Z$ exchange diagram can be derived
based on the interaction,
\begin{equation}
 {\cal L}_Z= - {g_2\over\cos\theta_W}Z
\left(  \bar f \gamma^\sigma (T^3_f - e_f\sin^2\theta_W) f
+ \tilde t^* {\cal N}_t
i{\stackrel{\leftrightarrow}{\partial^\sigma}} \tilde t
+ \tilde b^* {\cal N}_b
i{\stackrel{\leftrightarrow}{\partial^\sigma}} \tilde b
\right)   \ . \end{equation}
\begin{equation}  {\cal N}^q=
\left(\begin{array}{cc}
T^3_{q_L}\cos^2\theta_q -e_q\sin^2\theta_W & 
{1 \over 2}T^3_{q_L}\sin2\theta_q\\
{1 \over 2}T^3_{q_L}\sin2\theta_q  
& T^3_{q_L}\sin^2\theta_q -e_q\sin^2\theta_W
       \end{array}\right) \ .\end{equation}
$$\qquad\quad
\left( \frac{d_\ell}{e}\right)_{A^0Z} \ = -
\frac{3\alpha_{\rm em} \tan\beta 
({1\over2}T^3_{\ell_L} -e_\ell \sin^2\theta_W)}
     {32\pi^3\sin^2\theta_W\cos^2\theta_W}\, 
           \frac{m_\ell}{M^2_{A^0}}   $$
\begin{equation}\qquad\qquad\qquad\qquad\qquad\times
\sum_{q=t,b} \sum_{a,b}  {\cal N}^q_{a,b} e_q \hbox{  Re}(\xi_q)_{a,b} 
{\cal F}\left({M_Z^2\over M^2_{A^0}},
              {M^2_{\tilde{q}_a} \over M^2_{A^0}}, 
              {M^2_{\tilde{q}_b} \over M^2_{A^0}}\right) \ .
\end{equation}
Although we explicitly use the lepton $\ell$ for our EDM calculation,
the formalism can be easily extended to the quark. The color EDM of
quark can  also be obtained from Eq.~(\ref{eq:Agamma})
by changing color factors.  After including the renormalization group
evolution factor, we use the simple nonrelativistic quark model 
to estimate the neutron EDM\cite{ckp}.

In Fig.~3 we show contributions to the EDM of the electron and the
neutron.  For EDMs of the electron and the quarks, contributions from the
charged Higgs boson (plus $W$ boson) and from the pseudoscalar neutral
Higgs boson (plus either photon and $Z$ boson) are all shown for
comparison.  For the EDM of the neutron, there is additional contribution due
to chromoelectric dipole moments of quarks from two--loop diagrams with the
pseudoscalar neutral Higgs boson plus gluons.  For simplicity, we
assume the $\mu$ parameter is real and 
$M^2_{\tilde q_L}=M^2_{\tilde q_R}=0.6$~TeV,
$ A_t = A_b = i(1$~TeV) as in Ref.\cite{ckp}.  In Figs.~3a and 3b, we
display the $\tan \beta$ dependence of each contribution to the EDM for
the electron and the neutron respectively for both 
$M_A=150$~GeV and $300$~GeV ($\mu= 1$~TeV).
In Figs.~3c and 3d we display the $\mu$ dependence for the same choice
of parameters with $\tan\beta=20$.  
In the minimal supersymmetric Standard Model (MSSM), the
masses of the charged Higgs boson and the pseudoscalar boson 
are related by 
$$M^2_{H^{\pm}}= M^2_{A^0} + M^2_W + \epsilon \ ,$$ 
where $\epsilon = 3g_2^2m_t^4 \ln(1+(\tilde{m}^2/m_t^2)) /(8\pi^2 M_W^2)$ 
is the the quantum correction\cite{charged}.  We have used the tree
level result in plotting Figs.~3a--d.  However, if quantum corrections are
included, $M_{H^{\pm}}$ and $M_A$ become less closely related.  Therefore
in Figs.~3e and 3f we display the dependence on $M_{H^\pm}$ and the dependence
on $M_A$ for their respective contributions.  Overall, the EDMs are not
very sensitive functions of $M_{H^\pm}$ and $M_A$.  
For the EDM of the neutron, the
chromoelectric dipole moment contribution (due to pseudoscalar boson 
exchanges) still dominates over other sources.  For the EDM of the 
electron, the contribution of the charged Higgs boson exchange is in
general small than that of the pseudoscalar exchange by about an order of
magnitude.  The $A^0Z$ exchange diagrams, which have not been 
included in Ref.\cite{ckp} contribute  even smaller.  

\section*{Conclusion}
We have calculated the charged Higgs related two--loop contributions to the
electric dipole moments of electron and neutron.  For numerical 
simplicity, we ignore the generational mixing between squarks,
however, it should be quite easy to incorporate if necessary.  We find
that the charged Higgs contribution are generally smaller than the
neutral Higgs contribution calculated earlier\cite{ckp} in MSSM without
the squark flavor mixing which we assume.  However, one can imagine that
in theories beyond MSSM, the two contributions may involve independent
parameters that should be constrained separately.  It is
straightforward to generalize our analytic result
to accommodate theories beyond MSSM or to include squark flavor mixing.

While we were preparing this manuscript, 
we became aware of a preprint\cite{ref:pil}
which aims to calculate the same contribution.
Our analytic results in Eq.~(\ref{eq:Im}) and Eq.~(\ref{eq:edmW}) 
differ from those in Ref.\cite{ref:pil}. 
In particular our amplitude in Eq.~(\ref{eq:gWH}) has the gauge invariant form,
while the corresponding one in Ref.\cite{ref:pil} is not.  This difference 
does not affect the order of magnitude of the numerical results very much.
However, the result for the charged Higgs contribution to EDM of electron
in Ref.\cite{ref:pil} starts decreasing for large $\tan \beta (>35)$, but
not in our gauge invaraiant result as shown in Fig.~3a. 

\section*{Acknowledgment}
We thank A. Pilaftsis for a discussion. 
DC and WFC are supported partially by a grant from National Science Council of
Republic of China and WYK partially by a grant from US Department of
Energy.  
DC and WYK wish to thank KIAS (Seoul) and Seoul National University
for hospitality during the SSS-99 workshop  when this work was initiated.

\newpage
\begin{figure}
\begin{center}
\begin{tabular}[b]{cc}
  \epsfysize=4.0in
    \epsffile[30 75 770 950]{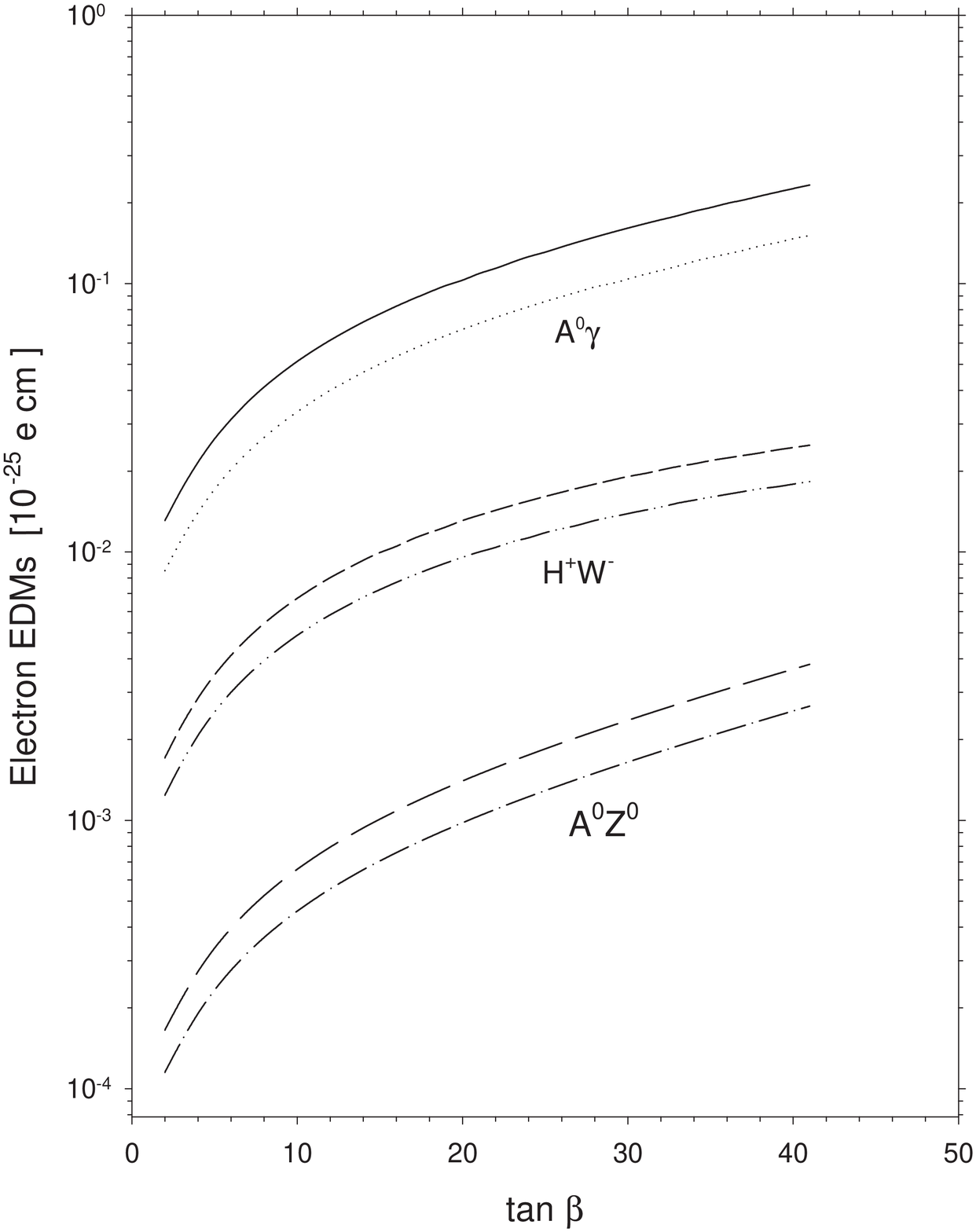} & 
  \epsfysize=3.8in
    \epsffile[30 75 770 950]{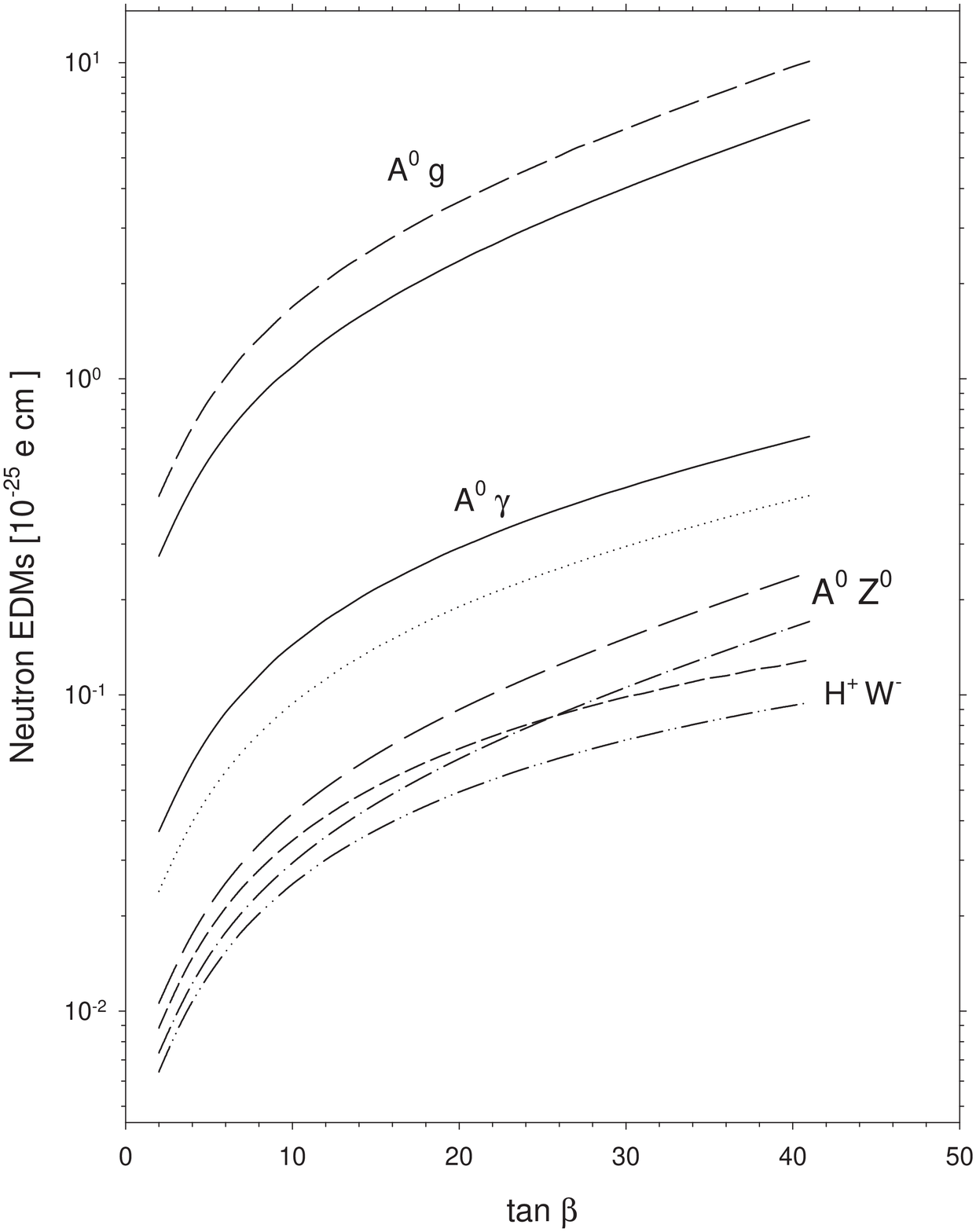}\\
       Fig.~3a & Fig.~3b               \\
\end{tabular}
\end{center}
\end{figure}
\begin{itemize}
\item[Fig.~3a\ ]
Numerical estimates of various contributions to the EDM of the electron as a
function of $\tan \beta$, for the case
$M^2_{\tilde q_L}=M^2_{\tilde q_R}=0.6$~TeV,
$M_A=150$~GeV and $300$~GeV,
real  $\mu=1$~TeV  and imaginary $ A_t = A_b = i(1$ ~TeV).
\item[Fig.~3b\ ]
Numerical estimates of various contributions to the EDM of the neutron as a
function of $\tan \beta$, for the case
$M^2_{\tilde q_L}=M^2_{\tilde q_R}=0.6$~TeV,
$M_A=150$~GeV and $300$~GeV,
real  $\mu=1$~TeV  and imaginary  $ A_t = A_b = i(1$ ~TeV).
\end{itemize}
\newpage

\begin{figure}
\begin{center}
\begin{tabular}[b]{cc}
  \epsfysize=4.0in
    \epsffile[30 75 770 950]{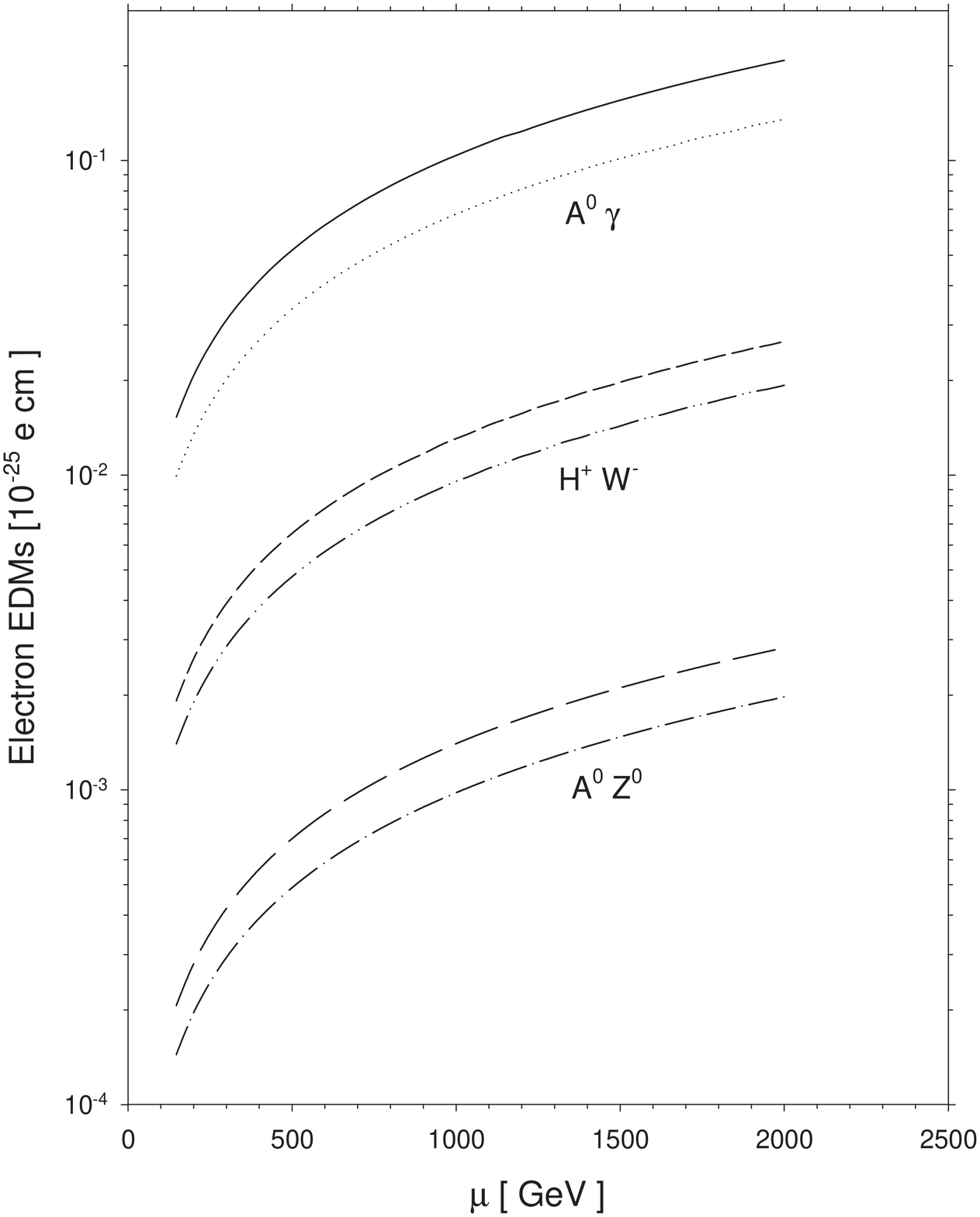} & 
  \epsfysize=3.8in
    \epsffile[30 75 770 950]{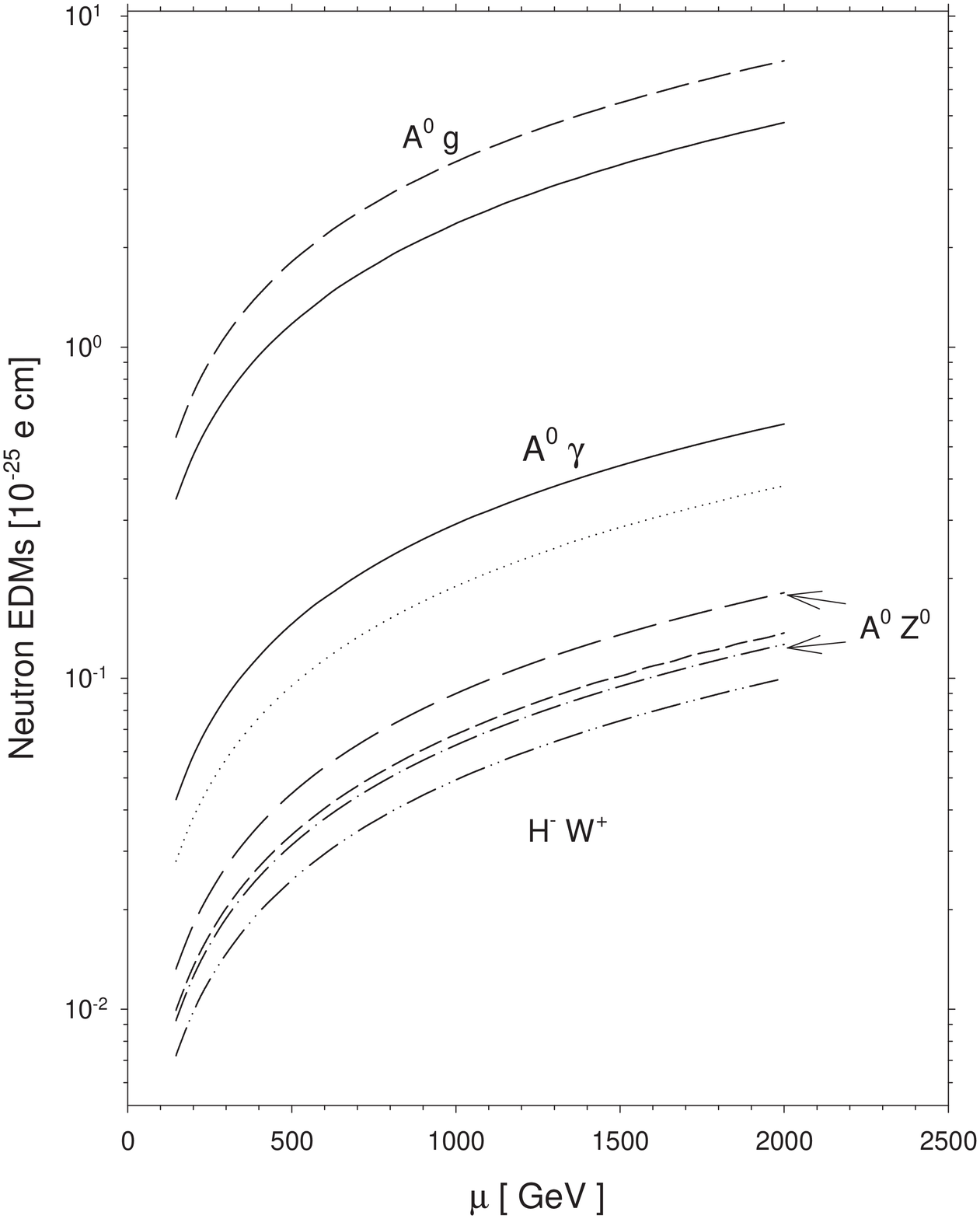}\\
       Fig.~3c & Fig.~3d               \\
\end{tabular}
\end{center}
\end{figure}
\begin{itemize}
\item[Fig.~3c\ ]
Numerical estimates of various contributions to the EDM of the electron as a
function of $\mu$, for the case $\tan\beta=20$,
$M^2_{\tilde q_L}=M^2_{\tilde q_R}=0.6$~TeV,
$M_A=150$~GeV and $300$~GeV,
real  $\mu$  and imaginary  $ A_t = A_b = i(1$ ~TeV).
\item[Fig.~3d\ ]
Numerical estimates of various contributions to the EDM of the neutron as a
function of $\mu$, for the case $\tan\beta=20$,
$M^2_{\tilde q_L}=M^2_{\tilde q_R}=0.6$~TeV,
$M_A=150$~GeV and $300$~GeV,
real  $\mu$  and imaginary  $ A_t = A_b = i(1$ ~TeV).
\end{itemize}
\newpage
\begin{figure}
\begin{center}
\begin{tabular}[b]{cc}
  \epsfysize=4.0in
    \epsffile[30 75 770 950]{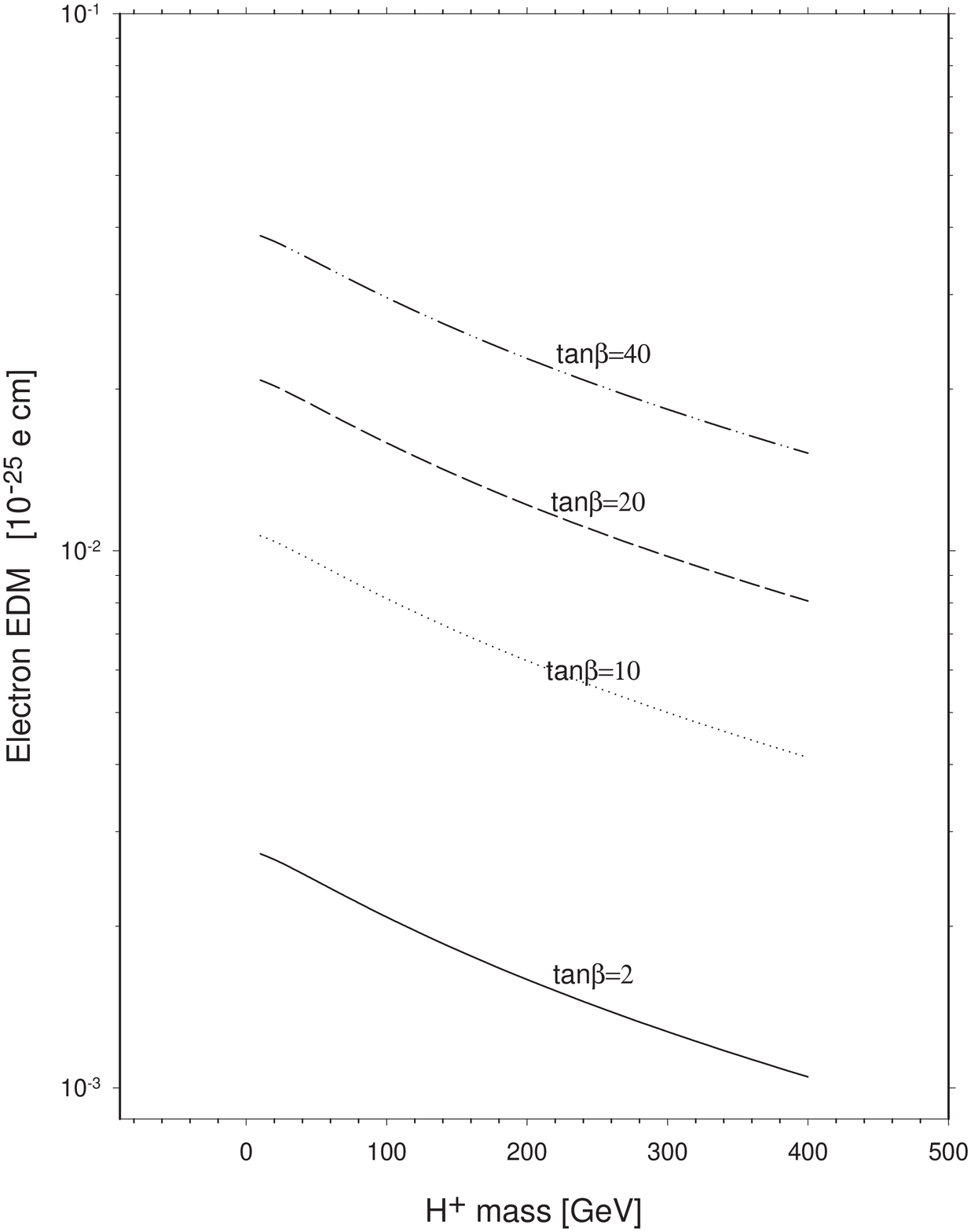} & 
  \epsfysize=3.8in
    \epsffile[30 75 770 950]{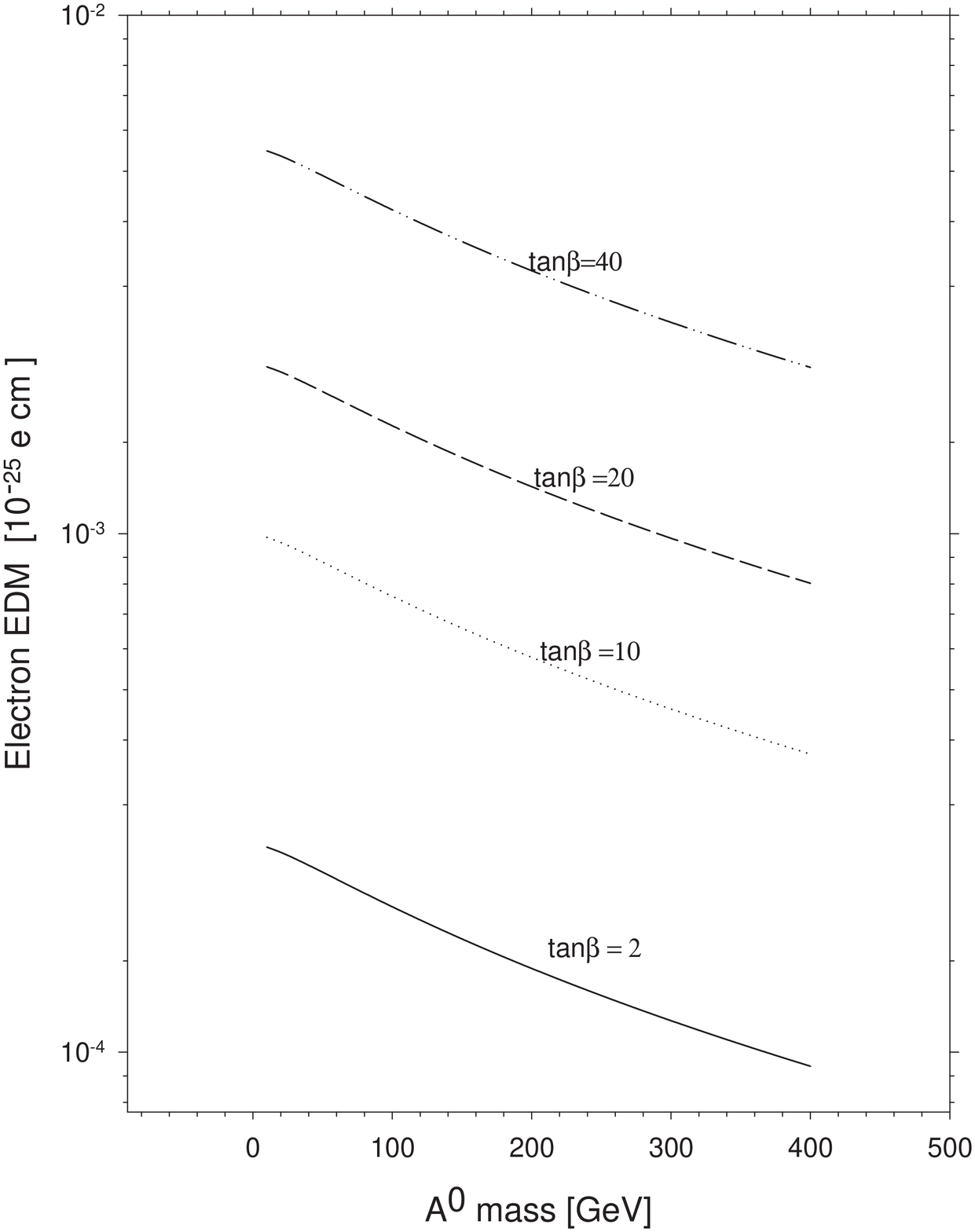}\\
       Fig.~3e & Fig.~3f               \\
\end{tabular}
\end{center}
\end{figure}
\begin{itemize}
\item[Fig.~3e\ ]
Numerical estimates of $H^+$ contributions to the EDM of the
electron as a
function of $M_{H^\pm}$, for the case
$M^2_{\tilde q_L}=M^2_{\tilde q_R}=0.6$~TeV,
real  $\mu=1 $~TeV  and imaginary  $ A_t = A_b = i(1$ ~TeV).

\item[Fig.~3f\ ]
Numerical estimates of $A^0$ contributions to the EDM of the
electron as a
function of $M_A$, for the case
$M^2_{\tilde q_L}=M^2_{\tilde q_R}=0.6$~TeV,
real  $\mu=1 $~TeV  and imaginary  $ A_t = A_b = i(1$ ~TeV).
\end{itemize}
\end{document}